\documentclass[prl,twocolumn,showpacs,superscriptaddress,nofootinbib]{revtex4-1}
\usepackage{amssymb}
\usepackage{amsbsy}
\usepackage{amsmath}
\usepackage[pdftex]{graphics}
\usepackage{graphics}
\usepackage{graphicx}
\usepackage[colorlinks,linkcolor=blue]{hyperref}

\def\la{\langle}
\def\ra{\rangle}

\def\tr{{\rm tr}}

\newcommand{\beq}{\begin{equation}}
\newcommand{\eeq}{\end{equation}}
\newcommand{\beqa}{\begin{eqnarray}}
\newcommand{\eeqa}{\end{eqnarray}}

\newcommand{\bra}[1]{\langle{#1}|}
\newcommand{\ket}[1]{|{#1}\rangle}
\usepackage{mathptmx} %comapct fonts

\begin{document}

%\title{Speed limits in open quantum system}
%\title{Speed limits in quantum open systems}
\title{Quantum speed limits in open system dynamics}

\author{A. del Campo}
\affiliation{Theoretical Division,  Los Alamos National Laboratory, Los Alamos, NM 87545, USA}
\affiliation{Center for Nonlinear Studies,  Los Alamos National Laboratory, Los Alamos, NM 87545, USA}
\author{I.~L. Egusquiza}
\affiliation{Department of Theoretical Physics and History of Science, UPV-EHU, 48080 Bilbao, Spain}
\author{M.~B. Plenio}
\affiliation{Institut f{\"u}r Theoretische Physik, Albert-Einstein Allee 11, Universit{\"a}t Ulm, D-89069 Ulm, Germany}
\affiliation{Institut for Integrated Quantum Science and Technology, Albert-Einstein Allee 11, Universit{\"a}t Ulm, D-89069 Ulm, Germany}
\author{S.~F. Huelga}
\affiliation{Institut f{\"u}r Theoretische Physik, Albert-Einstein Allee 11, Universit{\"a}t Ulm, D-89069 Ulm, Germany}
\affiliation{Institut for Integrated Quantum Science and Technology, Albert-Einstein Allee 11, Universit{\"a}t Ulm, D-89069 Ulm, Germany}

\begin{abstract}
Bounds to the speed of evolution of a quantum system are of fundamental interest
in quantum metrology, quantum chemical dynamics and quantum computation. We derive
a time-energy uncertainty relation for open quantum systems undergoing a general,
completely positive and trace preserving (CPT) evolution which provides a bound to
the quantum speed limit. When the evolution is of the Lindblad form, the bound is
analogous to the Mandelstam-Tamm relation which applies in the unitary case, with
the role of the Hamiltonian being played by the adjoint of the generator of the
dynamical semigroup. The utility of the new bound is exemplified in different scenarios,
ranging from the estimation of the passage time to the determination of precision
limits for quantum metrology in the presence of dephasing noise.
\end{abstract}

%03.65.-w: Quantum mechanics
%03.65.Yz 	Decoherence; open systems; quantum statistical methods (see also 03.67.Pp in quantum information; for decoherence in Bose-Einstein condensates, see 03.75.Gg)
%03.67.Lx	Quantum computation architectures and implementations

%03.65.Xp 	Tunneling, traversal time, quantum Zeno dynamics

\maketitle
\noindent

%\paragraph{Introduction}
How fast can a quantum system evolve? Quantum mechanics acts as a legislative body
imposing speed limits to the evolution of quantum systems. While these limits are
both ultimate and fundamental, at the same time, their existence is at the center
of a surge of activity, as a result of their manifold applications, including the
identification of precision bounds in quantum metrology \cite{GLM11}, the formulation
of computational limits of physical systems \cite{Lloyd00}, and the development of
quantum optimal control algorithms \cite{Caneva09}.

Bounds on the speed of evolution are intimately related to the concept of the passage
time $\tau_P$, which is the required time for a given pure state $\ket{\chi}$ to become
orthogonal to itself under unitary dynamics \cite{Schulman08}. One of the early answers
to this problem was provided by Mandelstam and Tamm (MT), who showed that the passage
time can be lower-bounded by the inverse of the variance in the energy of the system
so that
\beqa
    \tau\geq \frac{\pi}{2}\frac{\hbar}{\Delta H}, \label{eq1}
\eeqa
where $\Delta H=(\la H^2\ra-\la H\ra^2)^{1/2}$, whenever the dynamics under study is
governed by an Hermitian Hamiltonian $H$ \cite{MT45,Fleming73,AA90,Vaidman92,Uhlmann92,Uffink93,DL11,kok,Zwierz12}. A
simple geometric interpretation of this result was provided by Brody using the Fubini-Study
metric in the Hilbert space spanned by the initial state and its orthogonal complement
\cite{Brody03}.
Indeed, the passage time problem can be posed as a quantum brachistochrone
problem. From this perspective, a particularly exciting result was found: whenever the
Hamiltonian is non-Hermitian PT-symmetric, the passage time can be made arbitrarily
small without violating the time-energy uncertainty principle \cite{Bender07,Mostafazadeh07}.
A second bound, due to Margolus and Levitin (ML), takes the simpler form $\tau \geq
\frac{\pi}{2}\frac{\hbar}{\la H\ra-E_0}$ where the zero of energy is generally shifted
to the ground state energy so that $E_0=0$ \cite{ML98}. This bound has been applied to
ascertain fundamental computational limits in nature \cite{Lloyd00,LT09}.

Despite the growing body of literature on the subject, the analysis has almost exclusively
been focused on unitary dynamics of isolated quantum systems. An analogous bound for open
quantum systems is highly desirable, since ultimately all systems are coupled to an environment
\cite{breuer,rivas}. As an example, such a bound on the evolution of an open system would
help to address the robustness of quantum simulators and computers against decoherence
\cite{CZ12}, as well as the relevance of the specific nature of the noise, and in particular
whether or not it is Markovian, in phase estimation problems of interest in metrology and
precision spectroscopy \cite{Huelga97,nmm}.

The MT bound can be derived by considering the time evolution of the overlap
$\alpha = |\langle\psi_t|\psi_0\rangle|$ between the initial state $|\psi_0\rangle$
and the quantum state $|\psi_t\rangle$ at time $t$ subject to a unitary evolution
$U(t) = \exp\{-iHt/\hbar\}$. It can be shown that the MT-limit (eq. \ref{eq1}) is
achievable, as for a suitable Hamiltonian $H$ we can satisfy the differential
equation $\hbar\frac{d\alpha^2}{dt} = -2\Delta H \alpha \sqrt{1-\alpha^2}$
which for $\alpha=\cos\phi$ is easily seen to result in $\hbar\dot\phi=\Delta H$
thus matching the MT bound \cite{vaidman}.

In the case of open system dynamics we need to consider general non-unitary quantum
evolutions and have the freedom to choose a variety of distance measures between
quantum states. One natural choice here is the fidelity between two mixed states
$\rho$ and $\sigma$, which is given by $F(\rho,\sigma) = \tr[\sqrt{\sqrt{\rho}\,\sigma\sqrt{\rho}}]$.
The quantum speed limit then provides a lower bound on the time $\tau$ that is
required to achieve, for a given initial state $\rho(0)$ and a target fidelity
$f_{target}$, the condition $F(\rho_t,\rho_0) < f_{target}$ subject to an open
system evolution. Ideally, such bounds should reduce to the MT bound in the case
of unitary dynamics on pure states and/or be easy to compute.

Bounds on $\tau$ may be derived by taking inspiration from the variational characterization
of the fidelity $F(\rho^S,\sigma^S) = \max[|\langle \psi^{SE}|\phi^{SE}\rangle|]$ \cite{Jozsa},
where the maximization is over all $|\psi^{SE}\rangle$ ($|\phi^{SE}\rangle$) on a larger Hilbert
space ${\cal H}^{SE}$ that are purifications of the mixed states $\rho^S$ ($\sigma^S$) on the smaller
system S, that is $\tr_E[|\psi^{SE}\rangle\langle\psi^{SE}|]=\rho^S$
($\tr_E[|\phi^{SE}\rangle\langle\phi^{SE}|]=\sigma^S$). Then for any specific purification the
inequality $F(\rho^S,\sigma^S) \ge |\langle \psi^{SE}|\phi^{SE}\rangle|$ holds. A general time
evolution of a subsystem $\dot\rho^S = {\cal L}\rho^S$ can always be generated by a joint
unitary dynamics $U^{SE}_{t}$ of the system with an environment such that
$$\rho^S_t = \tr_B[U^{SE}_{t}|\psi^{SE}_{0}\rangle\langle\psi^{SE}_{0}| U^{\dagger,SE}_{t}]$$
where $\rho^{S}_{0} = \tr_B[|\psi^{SE}_{0}\rangle\langle\psi^{SE}_{0}|]$. Such a dynamics will
be generated by a suitable Hamiltonian $H^{SE}_t$ but it should be noted that the
choice of $|\psi^{SE}_{0}\rangle$, $U^{SE}_{t}$ and thus $H^{SE}_t$ is not unique.

Now we can make use of the fact that
$$f_{target} \ge F(\rho^{S}_{t},\rho^{S}_{0}) \ge F(U^{SE}_{t}|\psi^{SE}_{0}\rangle,|\psi^{SE}_{0}\rangle)$$
for any choice of purification of $\rho^{S}_{t}$ and $\rho^{S}_{0}$ and any choice of unitary
dynamics $U^{SE}_{t}$ that generates $\dot\rho^S = {\cal L}\rho^S$ on the subsystem. This
implies that any choice of purification and unitary evolution will achieve
$f_{target} \ge F(U^{SE}_{t}|\psi^{SE}_{0}\rangle,|\psi^{SE}_{0}\rangle)$ at an earlier time $t$
than $f_{target} \ge F(\rho^{S}_{\tau},\rho^{S}_{0})$, i.e. $t<\tau$. As a consequence, for any
choice of $|\psi^{SE}_{0}\rangle$, $U^{SE}_{t}$ and thus $H^{SE}_t$ we obtain a lower bound
on $\tau$. If $H^{SE}_t$ is given, then we can compute $\Delta H^{SE}_t$ in the state
$|\psi^{SE}_{t}\rangle$ and immediately provide a lower bound on $\tau$ via the MT and
$\hbar\Delta\phi = \hbar\int_0^t ds \dot\phi \le \int_0^t ds \Delta H^{SE}_s$.
Needless to say, performing the optimization over all possible purifications and all
possible $H^{SE}_t$ is a challenging task that will be very hard to perform in the general
case. Two routes are suggested themselves. Firstly, well chosen $|\psi^{SE}_{0}\rangle$, $U^{SE}_{t}$
and thus $H^{SE}_t$ will lead to excellent bounds for reasonably simple cases. Secondly,
analytical lower bounds on $\tau$ may also be obtained by studying different distance measures
that are easier to handle and thus admit closed formulae for lower bounds.

Here we follow this second approach to find an analytical and easy to compute lower bound
on the speed of evolution in open quantum systems. We shall derive a bound analogous to the
seminal result by MT where the energy variance of the initial state is replaced by a more
general measure taking into account the coupling to the environment. We shall pay particular
attention to the dynamics governed by a dynamical semigroup in which case the evolution of
the system is ruled by a master equation of the Lindblad form \cite{lindblad}. 
%It has recently
%been pointed out that the MT and ML bounds are violated under non-unitary dynamics \cite{kok}.
We shall  show in the following that Markovian systems are subjected to a MT-type
of bound where the adjoint of the generator of the dynamical semigroup plays the role of the
system Hamiltonian in the unitary case.

\paragraph{Decay of an  open quantum system}\label{sec-combined-system}

%To date, studies of the passage time have nearly exclusively focused on the unitary dynamics.

Consider a given system described by a state $\rho_0$ (from now we drop the upper index S for
convenience) coupled to an environment in a state $\rho^{E}_0$, and assume both system and
environment are weakly coupled such that the initial global state can be approximated by
$\rho_0 \otimes \rho^{E}_{0}$. Let the global reversible dynamics be governed by a unitary
evolution operator $U_t$. The reduced dynamics of the system is given by a one-parameter
family of dynamical maps $\rho \mapsto {\cal V}_t\rho:=\tr_E[U_t\rho_0\otimes\rho^{E}_0 U_t^\dag]$,
parameterized by the time variable $t\in\mathbb{R}^+$. Whenever the typical time scale of the
environment is much smaller than that of the system, one can assume a Markovian
dynamics. Under Markovian dynamics, such maps form a quantum dynamical semigroup
${\cal V}_{t+s}\rho = {\cal V}_t{\cal V}_s\rho$, $t,s>0$ (we assume that the open
system is not subjected to an external time-dependent field so that the generator
of the quantum dynamical semigroup is time independent). Any such map can be
represented by a Markovian master equation
\beqa
    \frac{d\rho_t}{dt}=\mathcal{L}\rho_t,
\eeqa
where the generator of ${\cal V}_t$ admits the Lindblad form \cite{lindblad}
\beqa
    \mathcal{L}\rho=-\frac{i}{\hbar}[H,\rho]+\sum_k\left(F_k\rho F_k^{\dag}
    -\frac{1}{2}\left\{F_k^{\dag} F_k,\rho\right\}\right),
\eeqa
such that $\mathcal{V}_t\rho_0=e^{t\!\mathcal{L}}\rho_0$. In such scenario we might
pose the following question: Which is the bound to the speed of evolution from an
initial state  $\rho_0$ under the action of a quantum dynamical semigroup $\mathcal{V}_t$?
To answer this question we introduce as a figure of merit the so called relative
purity \cite{LCW98}
\beqa
 f(t) = \frac{\tr\left[\rho_0 \rho_t\right]}{\tr(\rho_0^2)},
\label{eq-relative-purity-subsystem}
\eeqa
which is a generalization of the survival probability $\mathcal{S}(t)=|\la\chi|e^{-iHt/\hbar}|\chi\ra|^2$ often used for a pure state $|\chi\ra$ subject to a Hamiltonian $H$, and that has proved useful in studying quantum speed limits in the unitary case \cite{LT09}.

\paragraph{Derivation of the bound from the (Lindblad) master equation}

Let us now characterize the decay rate of the relative purity. Note that whenever the generator admits
a Lindblad form (i.e. for a Markovian quantum master equation),
\beqa
    \dot{f}(t)
    =\frac{\tr\left[\rho_0 \;\mathcal{L}\rho_t\right]}{\tr(\rho_0^2)}
    =\frac{\tr\left[\mathcal{L}^{\dag}\rho_0 \;\rho_t\right]}{\tr(\rho_0^2)}
\eeqa
where the adjoint of the generator of the dynamical map reads
\beqa
    \mathcal{L}^{\dag}\rho_0=\frac{i}{\hbar}[H,\rho_0]+\sum_k\left(F_k^{\dag}\rho_0F_k
    -\frac{1}{2}\left\{F_k^{\dag} F_k,\rho_0\right\}\right).
\eeqa
The rate of change of $f$ can then be bounded using the Cauchy-Schwarz inequality for operators,
$|\tr(A^\dag B)|^2\leq \tr(A^\dag A)\tr(B^\dag B)$.
Then
\beqa
\label{boundineq}
|\dot{f}(t)|
%&=&\tr(\rho_0\mathcal{L}\rho_t)/\tr\rho_0^2\nonumber\\
&\leq&
\sqrt{\tr[(\mathcal{L}^\dag\rho_0)^2]\tr[\rho_t^2]}/\tr\rho_0^2\nonumber\\
&\leq&
\sqrt{\tr[(\mathcal{L}^\dag\rho_0)^2]}/\tr\rho_0^2,
\label{thebound}
\eeqa
that is, by making reference exclusively to the initial state and the dynamical map.
Let us parametrize $f(t)=\cos\vartheta$ with $\vartheta\in[0,\pi/2]$. Upon integration between $\vartheta=0$ ($f(0)=1$) and a final $\vartheta=\theta$, the following bound to the required time of evolution is found
\beqa
    \tau_{\theta }\geq\frac{|\cos\theta-1| \tr\rho_0^2}{\sqrt{\tr[(\mathcal{L}^\dag\rho_0)^2]}}\geq
    \frac{4\theta^2\tr\rho_0^2}{\pi^2\sqrt{\tr[(\mathcal{L}^\dag\rho_0)^2]}}.
    \label{bound}
\eeqa
Here, $v=\sqrt{\tr[(\mathcal{L}^\dag\rho_0)^2]}$ provides an upper bound to the speed of evolution.
This generalizes the MT uncertainty relation for open quantum systems governed by a Markovian quantum master equation.
The generalization to a time-dependent Lindbladian $\mathcal{L}(t)$ is straightforward and reads
\beqa
\label{bound1}
\tau_{\theta }\geq \frac{4\theta^2\tr\rho_0^2}{\pi^2\overline{\sqrt{\tr[(\mathcal{L}^\dag\rho_0)^2]}}}.
\eeqa
where $\overline{X}=\tau_{\theta }^{-1}\int_0^{\tau_{\theta}}Xdt$.

\paragraph{Derivation of the bound using general quantum channels}

To remove the Markovian approximation, we note that any kind of time evolution of a
quantum state $\rho_0$ can be written in the form $\rho_t=\sum_{\alpha}K_{\alpha}(t,0)\rho_0K_{\alpha}^{\dag}(t,0)$.
In particular, $K_{\alpha}(t,0)$ is independent of $\rho_0$ if the dynamical map is induced
from an extended system with the initial condition $\rho^{SE}_0=\rho_0\otimes\rho^E(0)$. Then,
the dynamical map is said to be universal. Let such map govern the evolution and consider
\beqa
    f(t)=\tr[\rho_0\rho_t]=\sum_{\alpha}\tr[\rho_0 K_{\alpha}(t,0)\rho_0 K_{\alpha}^{\dag}(t,0)].
\eeqa
Parametrizing $f(t)=\cos\theta$, a bound can be derived
\beqa
\label{bound2}
\tau_{\theta }&\geq&\frac{2\theta^2}{\pi^2}\frac{\sqrt{\tr[\rho_0^2]}}{\overline{\sum_{\alpha}||K_{\alpha}(t,0)\rho_0\dot{K}_{\alpha}^{\dag}(t,0)||}}
\eeqa
where  $||A||=\sqrt{\tr(A^{\dag}A)}$ is the Hilbert-Schmidt norm of $A$. Details of
the derivation are provided in \cite{SM}.

\paragraph{Applications}
The bound to the speed of evolution presented above is the main result of this paper.
In the following we shall analyze some particular cases to illustrate its use, see too \cite{SM}.

{\it Passage time.} -
Under unitary time evolution, the passage time is the minimum time required for a time
evolving state $|\chi(t)\ra$ to become orthogonal to its initial value $|\chi(0)\ra$.
Let us consider a pure state such that $\tr\rho_0^2=1$ and let
$\mathcal{L}^\dag\rho=-\mathcal{L}\rho=i[H,\ket{\chi}\bra{\chi}]/\hbar$.
It follows from Eq. (\ref{bound1}), that
\beqa
    \tau_{\pi/2}\geq\frac{\hbar}{\sqrt{2}\Delta E}.
\eeqa
Alternatively, for $\alpha=1$, $K=\exp[-i(H-\bra{\chi} H \ket{\chi})t/\hbar]$,
$\tau_{\theta}\geq\frac{\hbar}{2\Delta E}$, a factor $1/\sqrt{2}$ smaller.
A similar reduction of the bound occurs for time-dependent Hamiltonians, in agreement with \cite{DL11}.
The usual definition of the passage time $\tau_P=\frac{\pi\hbar}{2\Delta E}$,
refers to the orthogonalization  measured by the fidelity, as stated above
\cite{MT45,Fleming73,Vaidman92,Schulman08}

{\it Non-Hermitian Hamiltonians.} -
Non-Hermitian Hamiltonians are ubiquitous in quantum physics and enjoy of a wide range of applications
from quantum optics \cite{PlenioKnight} to reactive scattering \cite{Moiseyev}. Their standard derivation
is based on Feshbach's partitioning theory, that allows to describe the effective dynamics of a quantum
system governed by a Hamiltonian $H$, when restricted to a given subspace associated with projector $P$
(with complement $Q$, such that $P+Q=1$, $P^2=P$, $Q^2=Q$). The effective Hamiltonian governing the dynamics
in the restricted subspace, $H_{\rm eff}=PHP+PHQ(E-QHQ)^{-1}QHP$, is generally non-Hermitian. Under
$H_{\rm eff}$ the density matrix $i\dot\rho=(H_{\rm eff}\rho-\rho H_{\rm eff}^{\dag})/\hbar$. Similarly, in open systems
under Markovian dynamics it is customary to split the generator of the dynamical map in two contributions
$\mathcal{L}_c$ and $\mathcal{D}$, i.e. $\mathcal{L}=\mathcal{L}_c+\mathcal{D}$. $\mathcal{L}_c$ describes
the coherent evolution associated with the non-Hermitian Hamiltonian $H_{\rm eff}=H-i\hbar\frac{1}{2}\sum_kF_k^{\dag}F_k$,
while the dissipator $\mathcal{D}\rho=\sum_kF_k\rho F_k^{\dag}$ is associated with spontaneous decay, and
it is a jump operator \cite{PlenioKnight}. More generally, let $H_{\rm eff}=H-i\Gamma$, where $H$ and $\Gamma$
are both Hermitian operators, so that $\mathcal{L}_c\rho=-i[H,\rho]/\hbar-\{\Gamma,\rho\}/\hbar$. Noting that upon
setting $\mathcal{D}\rho=0$, the bound to the speed  of evolution under non-Hermitian Hamiltonians still
holds, it follows from Eq. (\ref{thebound}) that
\beqa
\label{eqbound}
\tau_{\theta}&\geq& \frac{4\theta^2\tr\rho_0^2}{\pi^2\sqrt{\tr[(\mathcal{L}_c^{\dag}\rho)^2]}},\nonumber\\
\nonumber\\
&=&\frac{4\theta^2\hbar\tr\rho_0^2}{\pi^2\sqrt{\tr(-[H,\rho]^2+\{\Gamma,\rho\}^2-2i[H,\Gamma]\rho^2})},\nonumber\\
&=&\frac{4\theta^2\hbar}{\sqrt{2}\pi^2\sqrt{\Delta H^2+(\la\Gamma^2\ra+\la\Gamma\ra^2)-i\la[H,\Gamma]\ra}},
\eeqa
where the last line applies exclusively to pure states. 
%Interestingly, in this case it is possible to rewrite the bound simply as $\tau_{\theta}\geq\frac{4\theta^2\hbar}{\pi^2v}$ where $v=\tr[(\partial_t\sqrt{\rho_t})^2]$ \cite{BG12}.
Using Eq. (\ref{bound2}) with $\alpha=1$, $K=\exp[-i(H-\bra{\chi} H \ket{\chi})t/\hbar-(\Gamma-i\bra{\chi} \Gamma \ket{\chi})t/\hbar]$, one finds $1/\sqrt{2}$ times the same expression.
%{\it Disentangling and decoherence times \cite{DH04}}

\paragraph{From quantum speed limits to metrological bounds}
The ultimate bound to parameter estimation is dictated by the ability to efficiently
discriminate neighboring quantum states. In a seminal paper \cite{BC94}, Braunstein
and Caves (BC) derived a quantum Cramer-Rao bound for the uncertainty in the (local)
estimation of a classical parameter $\phi$ of the form:
\begin{equation}
    \Delta \phi \geq \frac{1}{\sqrt{\nu F_Q(\phi)}},
\end{equation}
where $F_Q$ denotes the quantum Fisher information and $\nu$ is the total number of
repetitions of the experiment where a $\phi$-dependence is linearly imprinted via a
general evolution. When the dynamics is unitary, an initial preparation of a probe
state in a {\em cat} (GHZ) state of $N$ subsystems allows to saturate the lower bound
and achieve a Heisenberg-limited resolution where $\Delta \phi \sim 1/N$. If the $N$
subsystems are used independently, so that the input state is factorizable as $N$
product states, only the standard scaling dictated by the central limit theorem $\Delta
\phi \sim 1/\sqrt{N}$ is achievable. This implies that the error bars in the actual
estimation of a parameter $\phi$ could be reduced by $1\sqrt{N}$ by means of employing
an entangled input probe provided that the system evolves unitarily. Whether or not
the standard scaling can be surpassed when the system's dynamics is open is
a most relevant issue where only partial results are known. Motivated by experiments
on precision spectroscopy, where a phase difference is estimated which is proportional
to the detuning between an external oscillator and a selected atomic frequency, we will
focus here on phase estimation problems under dephasing noise. Assuming decoherence to
be Markovian and affecting each subsystem independently (local noise assumption), it
was shown in \cite{Huelga97} that this type of noise renders product and maximally
entangled states metrologically equivalent, and argued that Markovian dephasing would
restore the standard scaling with an optimal resolution to be achieved by a type of
partially entangled states so that $\Delta \phi^{opt}/\Delta \phi^{p} = 1/\sqrt{e}$.
Subsequent work proved this bound to be achievable asymptotically \cite{kitagawa} but
only very recently it was proved in all generality that the bound is sharp and coincides
with the one imposed by the maximization of the quantum Fisher information \cite{Davidovich11}.
The metrological equivalence of product and maximally entangled state preparations under
Markovian decoherence can be predicted with the new bound eq. (\ref{bound}), which yields
the ratio $t_{\rm GHZ}=t_p/N$, where $t_{\rm GHZ}$ and $t_p$ are the optimal interrogation times
when using maximally entangled and product state inputs, respectively. This can be easily
shown by writing the dephasing master equation in the interaction picture as
\beqa
    \dot{\rho}=- \gamma \rho +\gamma\sigma_z\rho\sigma_z,
\eeqa
and considering a pure state $\rho_0=|\chi\ra\la\chi|$ of the form $|\chi\ra=(|0\ra+|1\ra)/\sqrt{2}$.
Then,
$\mathcal{L}^{\dag}\rho_0=-\gamma \rho_0 +\gamma\sigma_z\rho_0\sigma_z$  and
\beqa
    & &\tr[(\mathcal{L}^{\dag}\rho_0)^2]
    =
    2\gamma^2.
\eeqa

This yields a minimal orthogonalization time $t_p=1/\sqrt{2} \gamma$. Repeating the
same procedure for a maximally entangled input of the (GHZ) form
$\rho_0=|\chi\ra\la\chi|$ with $|\chi\ra=(|0\ra^{\otimes N}+|1\ra^{\otimes N})/\sqrt{2}$,
we obtain an optimal interrogation time $t_{\rm GHZ}=t_p/N$ which leads to
$\Delta \phi^{\rm GHZ}=\Delta \phi^p$ when the resolution is estimated operationally
as $\Delta \phi=<\Delta O^2>^2/\sqrt{\nu}\mid \frac{\partial <O>}{\partial \phi}\mid$,
with $O$ denoting a projective population measurement, which is known to be optimal
for this specific context. Alternatively, we can estimate the Fisher information in the form
\beqa
    F(\rho_{\phi})=\sum_i \frac{1}{p_i} \left(\frac{\partial p_i}{\partial \phi}\right)^2,
\eeqa
where $p_i=\tr(\rho_{\phi}P_i)$ and $P_i$ is a population projective measurement. Note
that this measurement procedure is optimal in this context. The resulting expressions
for product and cat states are, respectively
\beqa
    F_p &=& N e^{-2 \gamma t} \,t^2,\\
    F_{\rm GHZ} &=& N^2 e^{-2 N \gamma t} \,t^2.
\eeqa
The ratio $\Delta \phi^{\rm GHZ}/\Delta \phi^p=\sqrt{(\nu_p F_p)/(\nu_{ghz} F_{\rm GHZ})}$
therefore equals 1 when considering the optimal interrogations times as dictated by the
bound eq. (\ref{bound}). Moreover, for pure states $\rho_0=\ket{\chi}\bra{\chi}$ and the
case of Markovian pure dephasing $\mathcal{L}^\dag\rho_0 = \gamma\sum_k [- \rho_0 + \sigma_z^{(k)}\rho_0\sigma_z^{(k)}]$
we have that
$\sqrt{\tr[(\mathcal{L}^\dag\rho_0)^2]} = \gamma \sqrt{(N^2 - 2N \sum_k|\bra{\chi}\sigma_z^{(k)}\ket{\chi}|^2
+ \sum_{kl}|\bra{\chi}\sigma_z^{(k)}\sigma_z^{(l)}\ket{\chi}|^2} \le \sqrt{2}\gamma N$ 
(Note that this may be generalized to the mixed state case and any form of {\em local}
noise as the locality implies that number of terms in $\mathcal{L}^\dag\rho_0$
grows linearly in the number of subsystems $N$). Then
with eq. (\ref{bound}) and the fact that the Fisher information obeys $F\leq N^2$
\cite{smerzi1,smerzi2}, the limit on the speed of evolution imposes the persistence of the standard
scaling $\Delta \phi \sim 1/\sqrt{N}$ no matter how weak the dephasing rate. This is a result that
is now firmly established \cite{Davidovich11,rafal} and that comes out in a rather natural fashion
within this new framework.

So far we have exploited specifically the fact that the system's dynamics is ruled by a
Lindblad master equation. However, our general derivation considers a (linear) dynamical map that is trace preserving and completely positive (CPT) but not necessarily divisible \cite{markov}. 
As a result, the bound could be valid for non Markovian
dynamics as long as they admit a representation in terms of a CP map \cite{cpt}. 
We have evaluated the prediction
for the optimal interrogation times of product and cat states for a model of non-Markovian
dephasing of this type, as proposed in \cite{sonia}, and obtained the ratio $t_{\rm GHZ}=t_p/N$,
just as in the Markov case. This seems to be in contradiction with recent results for models
of non Markovian dephasing, which predict a ratio $t_{\rm GHZ}=t_p/\sqrt{N}$ \cite{nmm} and raises an interesting conjecture with which we finish this section. There could exist forms of coloured noise for which the metrological equivalence between cats-products input probes still holds. This inequivalence in the achievable resolution of a phase estimation could then be exploited to quantitatively quantify non-Markovianity.

\paragraph{Conclusions--}
A bound to the speed of evolution under an open-system dynamics has been provided, generalising
the classic result by  Mandelstam and Tamm known for the unitary case. In the Markovian limit,
we have shown that the adjoint of the generator of the dynamical semigroup plays the role of the
commutator with the Hamiltonian in the MT bound. Despite the fact that the bound is not tight,
in the sense of non coinciding with the unitary solution for closed systems, it allows to naturally
predict the inaccessibility of the Heisenberg limit under Markovian noise. Moreover, when using
the general form of the bound for universal channels, the new limit on the speed of evolution
suggest the inequivalence of different forms of coloured noise for precision spectroscopy.
Our results are applicable to a wide variety of scenarios including bounding decoherence
rates \cite{Zurek91}, and quantum speed limits in dissipative state preparation \cite{PlenioHuelga}, quantum computation and simulation assisted by dissipation \cite{frank}.

\paragraph{Note--} After the completion of this work,  we learned about reference \cite{Taddei12} devoted to quantum speed limits to the global unitary dynamics 
 of a system emmbedded in an environment.

\paragraph{Acknowledgements--}
It is a pleasure to thank D. Alonso, D.  J. Brody, B. Damski, M. Meister, A. Rivas, and A. Ruschhaupt for fruitful discussions and comments on the manuscript. This work was supported by the U.S. Department of Energy through the LANL/LDRD Program, the Basque Government (IT-559-10), the UPV/EHU UFI 11/55, the STREP PICC, the Alexander von Humboldt Foundation, the Integrated project QEssence and a  LANL J. Robert Oppenheimer fellowship.

%\vspace{0.21cm}

%\noindent
%{\bf Author Contribution:}\\
%\noindent
%AdC initiated the project and derived the bounds, ILE, MBP and SFG exemplified the implications of these bounds. 
%All authors contributed to writing the manuscript.

\section{Derivation of the bound using quantum channels}

Any kind of time evolution of a quantum state $\rho_0$ can be written in the form
$\rho_t=\sum_{\alpha}K_{\alpha}(t,0)\rho_0K_{\alpha}^{\dag}(t,0)$. In particular,
$K_{\alpha}(t,0)$ is independent of $\rho_0$ if the dynamical map is induced from
an extended system with the initial contidion $\rho^{SE}_0=\rho_0\otimes\rho^E_0$.
Then, the dynamical map is said to be universal. Let such map govern the evolution and
consider
\beqa
    f(t)=\frac{\tr[\rho_0\rho_t]}{\tr\rho_0^2} =
    \frac{1}{\tr\rho_0^2}\sum_{\alpha}\tr[\rho_0K_{\alpha}(t,0)\rho_0K_{\alpha}^{\dag}(t,0)].
\eeqa
For compactness, let us denote $K_{\alpha}=K_{\alpha}(t,0)$. It follows that
\beqa
\dot{f}(t)&=&\frac{\tr[\rho_0d_t\rho_t]}{\tr\rho_0^2},\nonumber\\
&=&\frac{1}{\tr\rho_0^2}\sum_{\alpha}\tr[\rho_0(\dot{K}_{\alpha}\rho_0K_{\alpha}^{\dag}+K_{\alpha}\rho_0\dot{K}_{\alpha}^{\dag}].
\eeqa
Taking the absolute value at both sides,
\beqa
|\dot{f}(t)|\leq\frac{1}{\tr[\rho_0^2]}\sum_{\alpha}\big[|\tr(\rho_0\dot{K}_{\alpha}\rho_0K_{\alpha}^{\dag})|+|\tr(\rho_0K_{\alpha}\rho_0\dot{K}_{\alpha}^{\dag})|\big].\nonumber\\
\eeqa
Using the Cauchy-Schwarz inequality $|\tr(AB)|\leq [\tr(A^{\dag}A)\tr(B^{\dag}B)]^{\frac{1}{2}}$,
\beqa
|\dot{f}(t)|
&\leq&\frac{1}{\sqrt{\tr[\rho_0^2]}}\sum_{\alpha}
(\sqrt{\tr[\dot{K}_{\alpha}\rho_0K_{\alpha}^{\dag}K_{\alpha}\rho_0\dot{K}_{\alpha}^{\dag}]}
\nonumber\\
& &+\sqrt{\tr[K_{\alpha}\rho_0\dot{K}_{\alpha}^{\dag}\dot{K}_{\alpha}\rho_0K_{\alpha}^{\dag}]}),
\nonumber\\
&=&\frac{2}{\sqrt{\tr[\rho_0^2]}}\sum_{\alpha}
\sqrt{\tr[\rho_0K_{\alpha}^{\dag}K_{\alpha}\rho_0\dot{K}_{\alpha}^{\dag}\dot{K}_{\alpha}]}.
\nonumber\\
\eeqa
We next use the fact that $|\int dt \dot{f}(t)|\leq\int dt |\dot{f}(t)|$.
Parametrizing $f(t)=\cos\theta$, a bound can be derived
\beqa
\tau
&\geq&\frac{|\cos\theta-1|}{2}\frac{\sqrt{\tr[\rho_0^2]}}{\overline{\sum_{\alpha}\sqrt{\tr[\rho_0K_{\alpha}^{\dag}K_{\alpha}\rho_0\dot{K}_{\alpha}^{\dag}\dot{K}_{\alpha}]}}},\nonumber\\\
&\geq&\frac{2\theta^2}{\pi^2}\frac{\sqrt{\tr[\rho_0^2]}}{\overline{\sum_{\alpha}\sqrt{\tr[\rho_0K_{\alpha}^{\dag}K_{\alpha}\rho_0\dot{K}_{\alpha}^{\dag}\dot{K}_{\alpha}]}}},
\nonumber\\\
&=&\frac{2\theta^2}{\pi^2}\frac{\sqrt{\tr[\rho_0^2]}}{\overline{\sum_{\alpha}||K_{\alpha}\rho_0\dot{K}_{\alpha}^{\dag}||}},
\eeqa
where  $\overline{X}=\tau_{\theta }^{-1}\int_0^{\tau_{\theta}}Xdt$ and $||A||=\sqrt{\tr(A^{\dag}A)}$ is the Hilbert-Schmidt norm of $A$.
 In the second line, we have used $|\cos\theta-1|\geq 4\theta^2/\pi^2$.

\section{Decoherence in an isotropic environment}
As an illustrative example, let us consider an exactly solvable model of open quantum dynamics.
The simplest model of decoherence in an isotropic environment is given
%by \cite{Diosi03}
\beqa
\label{mastereq}
d_t\rho=-\gamma\sum_i[\sigma_i,[\sigma_i,\rho]],
\eeqa
which is governed by a unital, relaxing dynamical map.
Consider the initial mixed state 
\beqa
\rho_0=\frac{1}{2}\left(
\begin{array}{cc}
1+r_3 & r_1-ir_2 \\
r_1+ir_2 & 1-r_3 \\
\end{array} \right),
\eeqa
with $r_i\in\mathbb{R}$ ($i=1,2,3$)
and purity $\tr\rho_0^2=\frac{1}{2}(1+\sum_ir_i^2)$. Its time evolution reads 
\beqa
\rho_t=\frac{1}{2}
\left(
\begin{array}{cc}
1+r_3s(t) & s(t)(r_1-ir_2) \\
s(t)(r_1+ir_2) & 1-s(t)r_3 \\
\end{array} \right),
\eeqa
where $s(t)=e^{-8\gamma t}\leq 1 \forall t\geq 0$, 
and asymptotically tends to
\beqa
\bar{\rho}=\frac{1}{2}\left(
\begin{array}{cc}
1 & 0 \\
0 & 1 \\
\end{array} \right).
\eeqa
The relative purity evolves according to
\beqa
f(t)=\frac{1+s(t)\sum_ir_i^2}{1+\sum_ir_i^2}.
\eeqa

The bound, written in the form of Eq. (7), reduces to 
\beqa
 |\dot{f}(t)|=\frac{8\gamma s(t)\sum_ir_i^2}{1+\sum_ir_i^2}
%\nonumber\\
\leq \frac{\sqrt{\tr[(\mathcal{L}^\dag\rho_0)^2]}}{\tr\rho_0^2}= \frac{8\sqrt{2}\gamma \sqrt{\sum_ir_i^2}}{1+\sum_ir_i^2},\nonumber\\
\eeqa
this is, $\sqrt{\sum_ir_i^2}s(t)\leq \sqrt{2}$ which holds $\forall t\geq 0$ given that $\sum_ir_i^2\leq 1$.
Bound Eq. (8) in the text reads
\beqa
\tau_{\theta }\geq\tau_B:=\frac{\theta^2\tr\rho_0^2}{\pi^2\gamma\sqrt{2\sum_ir_i^2}},
\eeqa
while the exact passage time  is given by
\beqa
\tau_\theta=-\frac{1}{8\gamma}{\rm log}\big[\frac{\cos\theta(1+\sum_ir_i^2)-1}{\sum_ir_i^2}\big].
\eeqa
We note that $f(\infty)=\frac{1}{2\tr{\rho_0^2}}$ so that for an initial pure state the maximum value of $\theta=\pi/3$.
%% ---------------- FIG. 1  ----------------
%
\begin{figure}
\begin{center}
\includegraphics[width=0.6\linewidth]{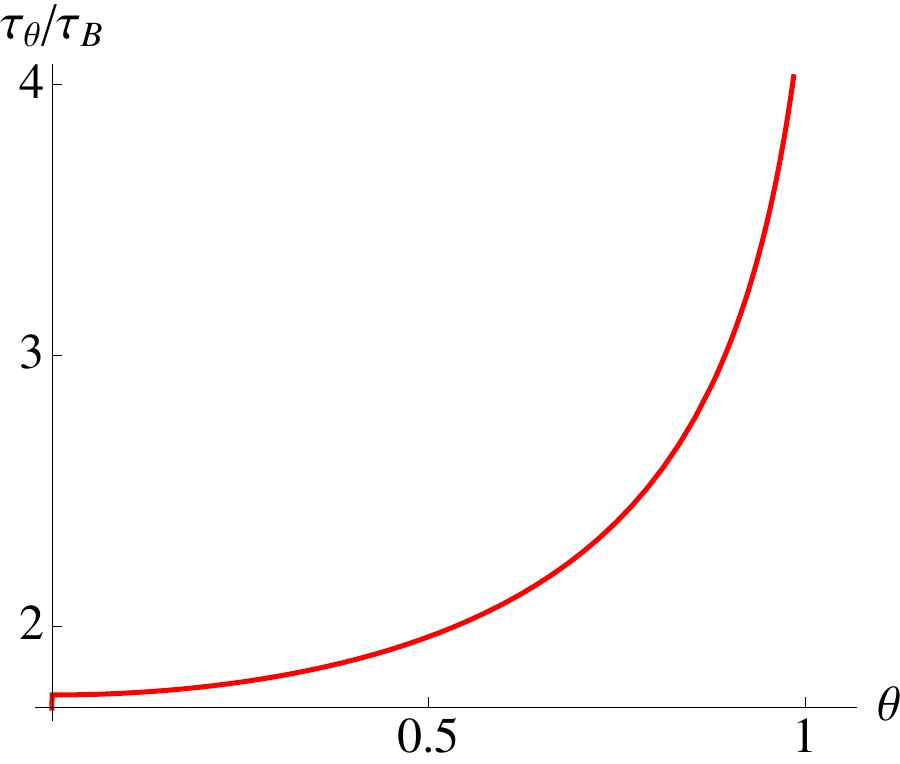}
\end{center}
\caption{\label{taup} Ratio between the exact passage time $\tau_\theta$ and its bound $\tau_B$ 
as a function of the angle $\theta$ parametrizing the relative purity, 
for an initial pute state placed in an isotropic environment.
}
\end{figure}
%------------------
Figure \ref{taup} shows the tighntness of the bound for an initial pure state undergoing decoherence according to  the master equation (\ref{mastereq}) as a function of the angle 
$\theta=\cos^{-1}f(\tau_{\theta})$.

\section{Dynamics in the presence of gain and loss}
Brody and Graefe have recently discussed the dynamics of a quantum
system in the presence of gains and losses of energy or amplitude
\cite{BG12}. In particular, they considered the master equation of the 
form 
\begin{eqnarray}
\label{gleq}
\hbar \, \partial_t\rho=-i[H,\rho]-\{\gamma,\rho\}+2{\rm tr}(\rho\Gamma)\rho.
\end{eqnarray}
In the case of a unitary dynamics for which $\Gamma=0$, the speed of 
evolution, defined by $v=\sqrt{ {\rm tr}[(\partial_t\sqrt{\rho})^2]}$, has been 
shown in \cite{Brody11} to be given by the 
skew-information measure of Wigner and Yanase \cite{WY63}. For 
$\Gamma\neq0$ they found that the speed of evolution, when $\rho$ is 
pure, to be given by the expression \cite{BG12} 
\begin{eqnarray}
v_t^2=2(\Delta H_t^2+\Delta\Gamma_t^2-i\langle [\Gamma,H]\rangle_t)/\hbar^2,
\end{eqnarray}
where all expectation values $\langle -\rangle_t$ are taken with respect
to the time-dependent solution $\rho_t$ of (\ref{gleq}). Since (\ref{gleq}) is 
nonlinear in $\rho$, the bounds derived in this paper is not necessarily 
applicable. However, it is interesting to note that for an initially pure state, 
Eq. (\ref{gleq}) is tantamount to a dynamical equation 
$\mathcal{L}\rho_0=\mathcal{L}_c\rho_0=-i[H,\rho_0]/\hbar-\{\Gamma-\langle 
\Gamma\rangle_0,\rho_0\}/\hbar$. 
This form allows us to establish a direct connection with the bound
Eq. (8) in the text,  by noting that the  velocity at $t=0$ is precisely
given by
\begin{eqnarray}
v_0^2 \equiv {\rm tr}[(\mathcal{L}^\dag\rho_0)^2].
\end{eqnarray}


\begin{thebibliography}{10}
\bibitem{GLM11} For a recent review, see V. Giovanetti, S. Lloyd, and L. Maccone,
{\em Advances in quantum metrology},
Nature Phot. {\bf 5}, 222-229 (2011).
%
\bibitem{Lloyd00}
S. Lloyd, Nature {\bf 406}, 1047 (2000);
S. Lloyd, Phys. Rev. Lett. {\bf 88}, 237901 (2002);
V. Giovannetti, S. Lloyd, and L. Maccone, Phys. Rev. A {\bf 67}, 052109  (2003).
 %
\bibitem{Caneva09}  T. Caneva, M. Murphy, T. Calarco, R. Fazio, S. Montangero, V. Giovannetti, and G. E. Santoro,
%Optimal control at the quantum speed limit.
Phys. Rev. Lett. {\bf 103}, 240501 (2009).
%
\bibitem{Schulman08} L. S. Schulman,
% Jump time and passage time: the duration of a quantum transition,
Lect. Notes Phys. {\bf 734}, 107 (2008).

\bibitem{MT45}  L.  Mandelstam and I. Tamm,
%The uncertainty relation between energy and time in nonrelativistic quantum mechanics.
J. Phys (USSR) {\bf 9}, 249 (1945).
For the subtleties associated with time-energy uncerttainty relations, see P. Busch, Lect. Notes Phys. {\bf 734}, 73 (2008).

\bibitem{Fleming73}  G. N. Fleming,
%A unitarity bound on the evolution of nonstationary states,
 Nuov. Cim. {\bf 16 A}, 232 (1973).
\bibitem{AA90} J. Anandan and Y. Aharonov, Phys. Rev. Lett. {\bf 65}, 1697 (1990).

\bibitem{Vaidman92}  L. Vaidman,  Am. J. Phys. {\bf 60}, 182 (1992).
\bibitem{Uhlmann92} A. Uhlmann Phys. Lett. A {\bf 161}, 329 (1992).

\bibitem{Uffink93}   J.  Uffink, Am. J. Phys. {\bf 61}, 935 (1993).
%
\bibitem{DL11}  S. Deffner and E. Lutz, arXiv:1104.5104.

\bibitem{kok} P.  J. Jones and P. Kok, Phys. Rev. A {\bf 82}, 022107 (2010).
\bibitem{Zwierz12} M. Zwierz Phys. Rev. A 86, 016101 (2012). 

\bibitem{Brody03}  D.~C. Brody, J. Phys. A: Math. Gen. {\bf 36}, 5587 (2003).


%
\bibitem{Bender07}  C. M. Bender, D. C.  Brody, H. F.  Jones,  and  B. K. Meister,
Phys. Rev. Lett. {\bf 98}, 040403 (2007).
%
\bibitem{Mostafazadeh07}  A. Mostafazadeh,  Phys. Rev. Lett. {\bf 99}, 130502 (2007).

\bibitem{ML98} N. Margolus and  L. B. Levitin,  Physica D {\bf 120}, 188 (1998).
%
\bibitem{LT09}  L. B. Levitin and  T. Toffoli,  Phys. Rev. Lett. {\bf 103}, 160502  (2009).
%
\bibitem{breuer} H. P. Breuer, F. Petruccione: {\it The Theory of Open Quantum Systems}, (Oxford University Press, New York 2002).
%
\bibitem{rivas} A. Rivas and S.~F. Huelga: {\it Open Quantum Systems. An Introduction}, (Springer, Heidelberg, 2011).
%
%
\bibitem{CZ12} J. I. Cirac and P. Zoller,
%Goals and opportunities in quantum simulation.
Nature Phys. {\bf 8}, 264 (2012).
%
\bibitem{Huelga97} S. F. Huelga,  C. Macchiavello, T. Pellizzari, A. K. Ekert,  M. B. Plenio, and J. I. Cirac,
%Improvement of frequency standards with quantum entanglement.
Phys. Rev. Lett. {\bf 79}, 3865 (1997).
%
\bibitem{nmm} A.~W. Chin, S.~F. Huelga, and M.~B. Plenio, %{\em Quantum metrology in non-Markovian environments},
Phys. Rev. Lett. {\bf 109}, 233601 (2012). 
%
\bibitem{vaidman}L. Vaidman,
%�Minimum time for the evolution to an orthogonal quantum state�,
Am. J. Phys. {\bf 60}, 182 (1992).
%
\bibitem{Jozsa} R. Jozsa, %Fidelity for mixed quantum states
J. Mod. Opt. {\bf 41}, 2315 (1994).
%
\bibitem{lindblad} A. Kossakowski, Bull. Acad. Polon. Sci. Math. {\bf 20}, 1021
(1972); V. Gorini, A. Kossakowski, and E. C. G. Sudarshan,
J. Math. Phys. {\bf 17}, 821 (1976); G. Lindblad,
Commun. Math. Phys. {\bf 48}, 119 (1976).
%




\bibitem{LCW98} K. M. R. Audenaert, E-print arXiv:1207.1197.
%D. A. Lidar,  I. L. Chuang, and K. B.  Whaley, Phys. Rev. Lett. {\bf 81}, 2590 (1998).

\bibitem{SM} See Supplemental Material at [URL will be inserted by publisher] for further applications of the bound. 

\bibitem{PlenioKnight} M. B. Plenio and P. L. Knight, Rev. Mod. Phys. {\bf 70}, 101 (1998).

\bibitem{Moiseyev} N. Moiseyev, Phys. Rep. {\bf 302}, 212 (1998).

%%%%%%%%%%%%%%%%%%%%%%%%%%%%%%%%%%%%%%%%%%%%%%%%%%%%%%%%%%%%%%%%%%%%%%%%%%%%%%%%%%%%%%%%%%%%%
% METROLOGY references

\bibitem{BC94} S. L. Braunstein and C. M. Caves,  Phys. Rev. Lett. {\bf 72}, 3439 (1994).
%
\bibitem{kitagawa} D. Ulam-Orgikh  and M. Kitagawa , Phys. Rev. A {\bf 64}, 052106 (2001).
%
\bibitem{Davidovich11} B. M. Escher, R. L. de Matos Filho, and L. Davidovich,
%General framework for estimating the ultimate precision limit in noisy quantum-enhanced metrology.
Nature Phys. {\bf 7}, 406-411 (2011).

\bibitem{smerzi1} L. Pezze and A. Smerzi, Phys. Rev. Lett. {\bf 102}, 100401 (2009).

\bibitem{smerzi2} 
 G. Toth, Phys. Rev. A {\bf 85}, 022322 (2012);
P. Hyllus, W. Laskowski, R. Krischek, C. Schwemmer, W. Wieczorek, H. Weinfurter,
L. Pezz\'e, and A.  Smerzi, Phys. Rev. A {\bf 85}, 022321 (2012).

\bibitem{rafal}  R. Demkowicz-Dobrzanski, J. Kolodynski, and M. Guta, %{\em The elusive Heisenberg limit in quantum enhanced metrology},
Nature Commun. {\bf 3}, 1063 (2012).

%
\bibitem{markov} A. Rivas, S. F. Huelga and M. B. Plenio, Phys. Rev. Lett. {\bf 105}, 050403 (2010); D. Chruscinski, A. Kossakowski, and A. Rivas, Phys. Rev. A {\bf 83}, 052128 (2011).  

\bibitem{cpt} The reliability of the requirement of CP has been discussed at length in the literature. See P. Pechukas, Phys. Rev. Lett. {\bf 73}, 1060 (1994) and reply by R. Alicki, Phys. Rev. Lett. {\bf 75}, 3020 (1995); For an example of a recent derivation of CP dynamics able to include non Markovian features, see A. Shabani and D. A. Lidar, Phys. Rev. A {\bf 71}, 020101 (R) 2005.

%
\bibitem{sonia} S. Daffer, K. W\'odkiewicz, J. D. Cresser, and J. K. McIver,  Phys. Rev. A {\bf 70}, 010304(R) (2004)

\bibitem{Zurek91} W. H. Zurek, Physics Today {\bf 44}, 36 (1991);  P. J. Dodd and  J. J. Halliwell, Phys. Rev. A {\bf 69}, 052105 (2004).

\bibitem{PlenioHuelga} M.~B. Plenio and S.~F. Huelga, Phys. Rev. Lett. {\bf 88}, 197901 (2002)

\bibitem{frank} F. Verstraete, M. Wolf, and J.~I. Cirac, Nature Physics {\bf 5}, 633 - 636 (2009).

\bibitem{Taddei12} M. M. Taddei, B. M. Escher, L. Davidovich, R. L. de Matos Filho, preceding Letter Phys. Rev. Lett. {\bf 110},
050402 (2013).
\end{thebibliography}

\begin{thebibliography}{10}
\bibitem{BG12} D. J. Brody and E.-M. Greaefe, Phys. Rev. Lett. {\bf 109}, 230405 (2012).
\bibitem{Brody11}D. J. Brody, J. Phys. A: Math. Gen. \textbf{44}, 252002 (2011).
\bibitem{WY63} E. P. Wigner and M. M. Yanase, Proc. Natl. Acad. Sci. {\bf 49}, 910 (1963). 
\end{thebibliography}
\end{document}